\documentclass[12pt]{article}
\usepackage{amsmath,amssymb} 
\usepackage{amsfonts}
\usepackage{amsthm}

\usepackage{titling}

\newtheorem{defn}{Definition}

\newtheorem{thm}{Theorem}[section]
\newtheorem{example}{Example}
\newtheorem{cor}[thm]{Corollary}
\newtheorem{prop}{Proposition}

\newtheorem{lem}[thm]{Lemma}
\newtheorem{conj}[thm]{Conjecture}
\newtheorem{constr}[thm]{Construction}
\newtheorem{note}{Note}

\newcommand{\bcor}{\begin{cor}}
	\newcommand{\ecor}{\end{cor}}
\newcommand{\bdefn}{\begin{defn}}
	\newcommand{\edefn}{\end{defn}}
\newcommand{\bnote}{\begin{note}}
	\newcommand{\enote}{\end{note}}
\newcommand{\bprop}{\begin{prop}}
	\newcommand{\eprop}{\end{prop}}
\newcommand{\blem}{\begin{lem}}
	\newcommand{\elem}{\end{lem}}
\newcommand{\bthm}{\begin{thm}}
	\newcommand{\ethm}{\end{thm}}
\newcommand{\bconj}{\begin{conj}}
	\newcommand{\econj}{\end{conj}}
\newcommand{\bconstr}{\begin{constr}}
	\newcommand{\econstr}{\end{constr}}

\newcommand{\bpf}{\begin{proof}}
	\newcommand{\epf}{\end{proof}}

\newcommand{\bfl}{\begin{flushleft}}
\newcommand{\efl}{\end{flushleft}}
\newcommand{\bit}{\begin{itemize}}
\newcommand{\eit}{\end{itemize}}
\newcommand{\ben}{\begin{enumerate}}
\newcommand{\een}{\end{enumerate}}
\newcommand{\bc}{\begin{center}}
\newcommand{\ec}{\end{center}}
\newcommand{\bfr}{\begin{flushright}}
\newcommand{\efr}{\end{flushright}}

\newcommand{\beq}{\begin{equation}}
\newcommand{\eeq}{\end{equation}}

\newcommand{\bea}{\begin{eqnarray}}
\newcommand{\eea}{\end{eqnarray}}
\newcommand{\bean}{\begin{eqnarray*}}
\newcommand{\eean}{\end{eqnarray*}}
\newcommand{\tr}{\ensuremath{\text{tr}}}

\newcommand{\calB}{\ensuremath{{\cal B}}}
\newcommand{\calA}{\ensuremath{{\cal A}}}
\newcommand{\calD}{\ensuremath{{\cal D}}}

\newcommand{\calDmod}{\ensuremath{{\cal D}_{\text{\tiny MOD}}}}

\newcommand{\mfdt}{\ensuremath{\text{MFD}_{2}}}
\newcommand{\izft}{\ensuremath{\text{IZ4}_2}}
\newcommand{\izfts}{\ensuremath{\text{IZ4}_{\text{$2$S}}}}

\newcommand{\dB}{\ensuremath{\text{dB}}}
\newcommand{\ffm}{\ensuremath{\mathbb{F}_{2^m}}}
\newcommand{\grm}{\ensuremath{GR(4,m)}}

\usepackage{scalerel,stackengine}
\stackMath
\newcommand\reallywidehat[1]{%
\savestack{\tmpbox}{\stretchto{%
  \scaleto{%
    \scalerel*[\widthof{\ensuremath{#1}}]{\kern-.6pt\bigwedge\kern-.6pt}%
    {\rule[-\textheight/2]{1ex}{\textheight}}
  }{\textheight}%
}{0.5ex}}%
\stackon[1pt]{#1}{\tmpbox}%
}
\begin{document}

\title{Quaternary and Component-Binary Spreading Codes with Low Correlation for Navigation Systems} 
\author{P. Vijay Kumar$^{*}$, Sugandh Mishra${\dagger}$, Dileep Dharmappa$^{\ddagger}$ \\ 
\normalsize  $^{*}$ Indian Institute of Science, Bengaluru, India\\
\normalsize  $^{\dagger}$ SAC, Indian Space Research Organization, Ahmedabad, India \\ 
\normalsize  $^{\ddagger}$ ISTRAC, Indian Space Research Organization, Bengaluru, India}
\date{December 3, 2024}
\maketitle

%
%


\begin{abstract}

In the first part of this two-part paper, we construct a family \mfdt\ of low correlation quaternary spreading codes having period $2046$.  By quaternary, we mean that the spreading code symbols are drawn from $Z_4$ and are designed to be used in conjunction with QPSK modulation. Apart from low auto and crosscorrelation properties, we also require in addition, to our knowledge for the first time, that the spreading code family \izft\  obtained by taking the union of the component in-phase and quadrature-phase binary spreading codes associated to each quaternary spreading code in \mfdt, also have desirable low correlation properties.  We also investigate the balance of the quaternary and binary spreading codes.  

The second part is motivated by an application to the design of spreading code, (in this application termed as ranging codes), having parameters suitable for use in a lunar PNT system.  Two lengths that are of particular interest for a planned lunar PNT satellite system are $2046$ and $10230$. We study the applicability of a subset \izfts\ of \izft\ containing balanced binary spreading codes having length $2046$ to such a lunar PNT system. We show that the spreading codes belonging to \izfts\ compare favorably with the spreading codes of length $2046$ appearing in a recent publication \cite{Dafesh2}. We also show that the IZ4$_{10}$ spreading code family in which the spreading codes have length $10230$, compares well in comparison with spreading codes of length $10230$ described in \cite{Dafesh2}.   In addition the IZ4$_{10}$ and IZ4$_2$ spreading codes have been paired so as to be orthogonal at zero shift despite their different lengths and chipping rates. 

\end{abstract} 

\section{Introduction} 

In the first part of this two-part paper, we construct a family \mfdt\ of low correlation quaternary spreading codes having period $2046$ .  By quaternary, we mean that the spreading code symbols are drawn from $Z_4$ and are designed to be used in conjunction with QPSK modulation. Apart from low auto and crosscorrelation properties, we also require in addition, to our knowledge for the first time, that the spreading code family \izft\  obtained by taking the union of the component in-phase and quadrature-phase binary spreading codes associated to each quaternary spreading code in \mfdt, also have desirably low correlation properties.  We also investigate the balance of the quaternary and binary spreading codes. 

The second part is motivated by an application to the design of spreading (also termed as ranging) codes having parameters suitable for use in a lunar PNT system.  Two lengths that are of particular interest for a planned lunar PNT satellite system are $2046$ and $10230$. We explore in the second part, the applicability of a subset \izfts\ of \izft\ containing balanced binary spreading codes having length $2046$ to such a lunar Position, Navigation and Timing (PNT) system. We show that the spreading codes belonging to \izfts\ compare favorably with the spreading codes of length $2046$ appearing in a recent publication \cite{Dafesh2}. We also show that the IZ4 spreading code family in which the spreading codes have length $10230$, compares well in comparison with spreading codes of length $10230$ described in \cite{Dafesh2}.  

\section{Mathematical Preliminaries}

Let
\bean
m \ = \  10, & & 
L \ = \  2^{m}-1 \ = \ 1023,
\eean
and $\alpha$ be the primitive element in $\mathbb{F}_{2^m}$ that is a zero of the primitive polynomial 
\bean
m_{\alpha}(x) & = & x^{10}+x^9+x^8+x^6+x^3+x^2+1.
\eean
The trace-dual basis $\{\delta_i\}_{i=0}^9$ associated to polynomial bases $\{\alpha^i \mid 0 \leq i \leq 9\}$ is given by 
\bean
\begin{array}{||c|c|c|c|c|c|c|c|c|c||} \hline \hline 
\delta_0 & \delta_1 & \delta_2 & \delta_3 & \delta_4 & \delta_5 & \delta_6 & \delta_7 & \delta_8 & 
\delta_9 \\ \hline \hline 
& & & & & & & & & \\ 
\alpha^{64} & \alpha^{63} & \alpha^{580} & \alpha^{138} & \alpha^{137} & \alpha^{136} & \alpha^{285} & \alpha^{284} & \alpha^{324} & \alpha^{65} \\ 
& & & & & & & & & \\ \hline \hline 
\end{array}
\eean
Let $\theta, \beta$ be defined by
\bean
\theta & = & \alpha^{65} \ + \ \alpha^{64}, \\ 
\beta & = & \alpha(1+2\theta). 
\eean
The it can be verified that $\tr(\theta)=1$.  
We will interchangeably use the symbol $\alpha$ to represent both the primitive element in \ffm\ having minimal polynomial 
\bean
m_{\alpha}(x) & = & x^{10}+x^9+x^8+x^6+x^3+x^2+1, 
\eean
as well as the unique element $\nu$ in the Teichmuller set of the Galois ring \grm\ having the property that 
\bean
\alpha & = & \nu \pmod{2}. 
\eean
This element $\nu$ has minimum polynomial over $Z_4$ given via Graeffe lifting (see \cite{KumDhaMis_TIT})
\bean
m_{\nu}(x) & = & x^{10}+x^9+3x^8+2x^7+x^6+x^3+x^2+2x+1. 
\eean

\section{Closed-Form Expression for Family \calD\ Sequence Family } 

\begin{defn} [Quaternary Family ${\cal D}$] \cite{TanUda} 
Let $m=10$, $\alpha$ be a primitive element of $\ffm$ and set $\beta=\alpha(1+2\theta)$ for some $\theta \in \ffm$, satisfying $\tr(\theta)=1$.   Let $H \subset \ffm$ be a maximal subset of \ffm\ satisfying 
\bean
y \in H & \Rightarrow & (y+\theta) \not \in H. 
\eean
Then the quaternary Family \calD~\cite{TanUda} is given by\footnote{The description in \cite{TanUda} is presented in terms of expressions for the binary components of the quaternary sequence family.}
\bean
\calD & = & \left\{  x 3^t \ + \ T([1+2y]\beta^t) \ \mid \ x \in \{0,1\}, \ \ y \in H \right\}. 
\eean
Then Family \calD\ contains $1024$ quaternary sequences of length $2046$. . 
\end{defn}
Let $\{Q_1(t)\}$ and $\{Q_2(t)\}$ be two sequences belonging to Family \calD\ given by 
\bea
Q_i(t) & = & x_i 3^t \ + \ T([1+2y_i]\beta^t), \ x_i \in \{0,1\}, \ \ y_i \in H, \ \ i=1,2.
\eea
Then the cross-correlation $\phi_{12}(\tau)$ of $\{Q_1(t)\}$ and $\{Q_2(t)\}$ is given by: 
\bean
\phi_{12}(\tau) & = & \sum_{t=0}^{2L-1} \imath^{Q_1(t+\tau)-Q_2(t)}.
\eean

\begin{thm} \label{thm:closed_form} We have the following closed-form expression for the correlation of two Family \calD\ PRN sequences associated to $y_1,y_2 \in H$ in the case of $\tau$ even:
\bean
\imath^{x_2-x_1}  \phi_{12}(\tau) & = & \bigg( -1 \ - \ (-1)^{(x_1-x_2)} \bigg) \ - \ \\ 
& & 32 \imath^{1-T(e(y_1,y_2,\tau))} \left\{1 \ - \ 
(-1)^{(x_1-x_2)} \bigg(\imath \times (-1)^{tr(e(y_1,y_2,\tau)\theta)}\bigg)\right\}.
\eean
When $\tau$ is odd, we have 
\bean
\imath^{x_2-3x_1}  \phi_{12}(\tau) & = & \bigg( -1 \ - \ (-1)^{(x_1-x_2)} \bigg) \ - \ \\ 
& & 32 \imath^{1-T(e(y_3,y_2,\tau))} \left\{ 1 \ - \ 
(-1)^{(x_1-x_2)} \bigg(\imath \times (-1)^{tr(e(y_3,y_2,\tau)\theta)}\bigg)\right\}, 
\eean
where $y_3=y_1+\theta$. 
\end{thm}
\begin{proof}
(see Section~\ref{sec:closed_form} for the proof) 
\end{proof}

\begin{cor}
It follows that $\phi_{12}(\tau)$ satisfies: 
\bean
 \phi_{12}(\tau) & \in & \left\{ \pm 32(1 \pm \imath), \ \ -2 \pm 32(1 \pm \imath)\right\} 
\eean
hence the maximum correlation magnitude satisfies, $\phi_{\max} \ \leq \ 46.69$. 
\end{cor}

\section{Component Binary Sequence Correlation} 

\subsection{Relating Quaternary and Binary Signals} 
Given a quaternary sequence $\{Q(t)\}$ let the binary sequences $\{u(t)\}$, $\{v(t)\}$ be defined by  
\bean
Q(t) & = & u(t) \ + \ 2 v(t), \ \ u(t), \ v(t) \in \{0,1\}, \\
w(t) & = & u(t) \ + \ v(t) \ \pmod{2}. 
\eean
Set 
\bean
\delta \ = \ \frac{1+\imath}{2}, & & 
\delta^* \ = \  \frac{1-\imath}{2}, \\ 
C(t) \ = \ \imath^{Q(t)} & = & \delta (-1)^{v(t)} \ + \ \delta^* (-1)^{w(t)}, \\
& = & \delta A(t) \ + \ \delta^* B(t).
\eean
We will refer to the sequences $\{w(t)\}$, $\{v(t)\}$ as the in-phase and quadrature-phase component sequences of the quaternary sequence $\{Q(t)\}$ respectively. 
This leads to the matrix equations:
\bean
\left[ \begin{array}{c} C(t) \\ C^*(t) \end{array} \right] 
& = & \underbrace{\left[ \begin{array}{cc} \delta & \delta^* \\
\delta^* & \delta \end{array} \right] }_{U}
\left[ \begin{array}{c} A(t) \\ B(t) \end{array} \right] , \\ 
UU^{\dagger} & = & I_2, \\
\left[ \begin{array}{c} A(t) \\ B(t) \end{array} \right]  
& = & \left[ \begin{array}{cc} \delta^* & \delta \\
\delta & \delta^* \end{array} \right]
\left[ \begin{array}{c} C(t) \\ C^*(t) \end{array} \right] , \\ 
\eean
It follows that 
\bean
(-1)^{v(t)} & = & \delta^* C(t) \ + \ \delta C^*(t) \ = \ \delta^* \imath^{Q(t)}  \ + \ \delta \imath^{-Q(t)}, \\
(-1)^{v(t)} & = & \delta C(t) \ + \ \delta^* C^*(t) \ = \ \delta \imath^{Q(t)}  \ + \ \delta^* \imath^{-Q(t)}, \\
\eean
Let
\bean
\rho_v(\tau) & = & \sum_{t=0}^{2L-1} (-1)^{v_1(t+\tau)-v_2(t)}, \\
\rho_w(\tau) & = & \sum_{t=0}^{2L-1} (-1)^{w_1(t+\tau)-w_2(t)}, \\
\rho_{v,w}(\tau) & = & \sum_{t=0}^{2L-1} (-1)^{v_1(t+\tau)-w_2(t)}.
\eean
Then we have 
\begin{thm}
\bean
\rho_v(\tau) & = & \Re\big(\phi_{12}(\tau)\big) \ + \ \Im\big(\Delta_{12}(\tau)\big), \\ 
\rho_w(\tau) & = & \Re\big(\phi_{12}(\tau)\big) \ - \ \Im\big(\Delta_{12}(\tau)\big), \\ 
\rho_{vw}(\tau) & = & \Re\big(\Delta_{12}(\tau)\big) \ + \ \Im\big(\phi_{12}(\tau)\big).
\eean
where
\bean
\phi_{12}(\tau) & = & \sum_{t=0}^{2L-1} C_1(t+\tau)C^*_2(t), \ \ \ \text{(Family ${\cal D}$ correlation)}\\
\Delta_{12}(\tau) & = & \sum_{t=0}^{2L-1} C_1(t+\tau)C_2(t), \ \ \ \text{(we will term this as a Family \calD\ anti-correlation)}. 
\eean
\end{thm}

\begin{proof}
(see Section~\ref{sec:bcorr} ) 
\end{proof}

\subsection{Computing the Anti-Correlation term $\Delta_{12}(\tau)$} 

We have that 
\bean
\Delta_{12}(\tau) & = & \sum_{t=0}^{2L-1} \imath^{Q_1(t+\tau)+Q_2(t)}. 
\eean
The only difference is that we have 
\bean
C_2(t) \ = \ \imath^{Q_2(t)} & \text{ in place of } & C^*_2(t) \ = \ \imath^{-Q_2(t)}.
\eean
Since
\bean
Q_2(t) & = & x_2 \ + \ T([1+2y_2]\beta^t), \\
\eean
we can simply replace 
\bean
x_2 \text{ by } 3x_2 & \text{ and } & y_2 \text{ by } (y_2+1) 
\eean
in the expressions for $\phi_{12}(\tau)$ to obtain the expressions for $\Delta_{12}(\tau)$.

\subsection{Correlation Expressions for $\Delta_{12}(\tau)$ in Summary}
We have in summary:
\bean
\Delta_{12}(\tau) 
& = & \begin{cases} \imath^{(x_1 -3x_2)} \bigg( \psi(y_1,y_2+1,\tau)  \ + \  (-1)^{(x_1 -3x_2)}  \psi(y_3,y_4+1,\tau)\bigg), \text{ for $\tau$ even }, \\ 
\imath^{(3x_1 -3x_2)} \bigg( \psi(y_3,y_2+1,\tau)  \ + \  (-1)^{(x_1 -3x_2)}  \psi(y_1,y_4+1,\tau)\bigg), \text{ for $\tau$ odd },
\end{cases} , \\
& = & \begin{cases} \imath^{(x_1 -3x_2)} \bigg( \psi(y_1,y_6,\tau)  \ + \  (-1)^{(x_1 -3x_2)}  \psi(y_3,y_8,\tau)\bigg), \text{ for $\tau$ even }, \\ 
\imath^{(3x_1 -3x_2)} \bigg( \psi(y_3,y_6,\tau)  \ + \  (-1)^{(x_1 -3x_2)}  \psi(y_1,y_8,\tau)\bigg), \text{ for $\tau$ odd },
\end{cases}  
\eean
where
\bean
\begin{array}{||c|c|c|c||} \hline \hline 
y_1 & y_3 & y_5 & y_7 \\ 
 & =(y_1+\theta) & =(y_1 + 1) & =(y_1 + \theta+1) \\ \hline \hline 
y_2 & y_4 & y_6 & y_8 \\ 
 & =(y_2+\theta) & =(y_2 + 1) & =(y_2 + \theta+1) \\ \hline 
%
%
\end{array} 
\eean
\bean
\psi(y_1,y_2,\tau) & := & \sum^{L-1}_{t=0} \imath^{T([1+2y_1]\alpha^{\tau}\alpha^t) \ - \ T([1+2y_2]\alpha^t)}, \\
\psi(y_3,y_4,\tau) & := & \sum_{t=0}^{L-1}  \imath^{T([1+2y_3]\alpha^{\tau}\alpha^t) \ - \ T([1+2y_4]\alpha^t)}.
\eean
etc.
Thus in summary, 
\bean
\Delta_{12}(\tau) 
& = & \begin{cases} \imath^{(x_1 -3x_2)} \bigg( \psi(y_1,y_6,\tau)  \ + \  (-1)^{(x_1 -3x_2)}  \psi(y_3,y_8,\tau)\bigg), \text{ even }, \\ 
\imath^{(3x_1 -3x_2)} \bigg( \psi(y_3,y_6,\tau)  \ + \  (-1)^{(x_1 -3x_2)}  \psi(y_1,y_8,\tau)\bigg), \text{ odd.} 
\end{cases}  
\eean

\subsection{Can $|\Delta_{12}(\tau)|$ be Large of $O(L)$ ?}

From \eqref{eq:Bone_qc}, we see that this can happen iff one of the following holds: 
\bean
y_1=y_6 \ \ \Leftrightarrow \ y_1=y_2+1 & \text{ and } & \tau \ = 0,  \\
y_3=y_8 \ \  \Leftrightarrow \ y_1=y_2+1 & \text{ and } & \tau \ =\ 0,   \\
y_3=y_6 \ \ \Leftrightarrow \ y_2=y_1+\theta+1  & \text{ and } & \tau \ = \ L, \\ 
y_1=y_8 \ \ \Leftrightarrow \ y_2=y_1+\theta+1  & \text{ and }  & \tau \ = \ L. 
\eean

We safeguard all of these settings by selecting $y$ from a subset $J$ of $H$ and ensuring that $y \in J \Rightarrow (y+1) \not \in J, \ (y+\theta+1) \not \in J$.

\begin{note} [Ensuring Good Quaternary Even Correlation]
Thus we note that even to ensure good quaternary anti-correlation, we need to ensure that 
\bean
y \in J & \Rightarrow & (y+1) \not \in J, \ \ (y+\theta) \not \in J, \ \ \text{ and } (y+\theta+1) \not \in J. 
\eean
\end{note}

\begin{defn} Let $J \subseteq \ffm$ be such that \bean
y \in J & \Rightarrow & \begin{cases} (y+\theta) \not \in J, \\  (y+1) \not \in J, \\  (y+\theta+1) \not \in J. \end{cases} 
\eean
Let the modification $\calDmod$ of Family \calD\ be defined by 
\bean
\calDmod & = & \left\{ \{x3^t \ + \ T([1+2y] \beta^t)\} \mid y \in J, x \in \{0,1\} \right\}. 
\eean
Note that this family has size $|\calB | \ = \ (2 \times 256) \ = \ 512$. 
\end{defn}

\begin{example}
We select as an example, 
\bean
\theta & =   \alpha^{64} \ + \ \alpha^{65}, \\ 
\eean
and it then turns out that 
\bean
T(\theta) \ = \  1, \ \Rightarrow \ \tr(\theta) \ = \  1. 
\eean
Let $H$ be the subset of \ffm\ given by 
\bean
H & = & \left\{ y=\sum_{i=0}^9 h_i \delta_i \mid h_8=h_9=0\right\}. 
\eean
The expansions with respect to basis $\{\delta_i\}_{i=0}^9$ of $1$ and $\theta$ are given respectively by 
\bean
1 & = & \delta_1+\delta_2+\delta_4+\delta_7+\delta_8, \\
\theta & = & \alpha^{64}+\alpha^{65} \ = \ \delta_0+\delta_9. 
\eean
It is clear that $H$ satisfies the requirement that 
\bean
y \in H & \Rightarrow & \begin{cases} y+1 \not \in H, & \\ y+\theta \not \in H, & \\ y+(\theta+1) \not \in H. &  \end{cases}
\eean
\end{example}
\begin{cor}
Arguing as in the case of $\{\phi_{12}(\tau)\}$, we can show that $\Delta_{12}(\tau)$ satisfies: 
\bean
\Delta_{12}(\tau) & \in & \left\{ \pm 32(1 \pm \imath), \ \ -2 \pm 32(1 \pm \imath)\right\} 
\eean
hence the maximum anti-correlation magnitude satisfies, $\Delta_{\max} \ \leq \ 46.69$. 
\end{cor}

\section{Summary Expressions for Quaternary and Binary Correlation} 

\subsection{Quaternary Correlation Summary}

\bean
\phi_{12}(\tau) 
& = & \begin{cases} \imath^{(x_1 -x_2)} \bigg( \psi(y_1,y_2,\tau)  \ + \  (-1)^{(x_1 -x_2)}  \psi(y_3,y_4,\tau)\bigg), \text{ even }, \\ 
\imath^{(3x_1 -x_2)} \bigg( \psi(y_3,y_2,\tau)  \ + \  (-1)^{(x_1 -x_2)}  \psi(y_1,y_4,\tau)\bigg), \text{ odd },
\end{cases}  
\eean
and 
\bean
\Delta_{12}(\tau) 
& = & \begin{cases} \imath^{(x_1 -3x_2)} \bigg( \psi(y_1,y_6,\tau)  \ + \  (-1)^{(x_1 -3x_2)}  \psi(y_3,y_8,\tau)\bigg), \text{ even }, \\ 
\imath^{(3x_1 -3x_2)} \bigg( \psi(y_3,y_6,\tau)  \ + \  (-1)^{(x_1 -3x_2)}  \psi(y_1,y_8,\tau)\bigg), \text{ odd.} 
\end{cases}  
\eean

\subsection{Binary Correlation Summary}

\begin{thm}
\bean
\rho_v(\tau) & = & \Re\big(\phi_{12}(\tau)\big) \ + \ \Im\big(\Delta_{12}(\tau)\big), \\ 
\rho_w(\tau) & = & \Re\big(\phi_{12}(\tau)\big) \ - \ \Im\big(\Delta_{12}(\tau)\big), \\ 
\rho_{vw}(\tau) & = & \Re\big(\Delta_{12}(\tau)\big) \ + \ \Im\big(\phi_{12}(\tau)\big).
\eean
\end{thm}
\begin{cor}
It follows from the set of possible values that can be taken on by $\{\phi_{12}(\tau)\}$, and the anti-correlation term $\Delta_{12}(\tau)$ that: 
\bean
\mid \rho_v(\tau)| & \leq & 66, \\ 
\mid \rho_w(\tau)| & \leq & 66, \\ 
\mid \rho_{vw}(\tau)| & \leq & 66. 
\eean
\end{cor}

\begin{defn}
Let
\bean
P & = & \left\{ (x,y) \mid x \in \{0,1\}, \ y \in J \right\},
\eean
and \izft\ be the family of binary sequences obtained by taking the union of the in-phase (IP) and quadrature-phase (QP) binary sequences, i.e., if 
\footnotesize  
\bean 
\text{QP}-\izft& = & \left\{ \{v(x,yt)\} \mid Q(x,y,t) = u(x,y,t)+2v(x,y,t), (x \in \{0,1\}, y \in J)x,y)\in P \right\},  \\
\text{IP}-\izft & = & \left\{ \{w(x,yt)\} \mid Q(x,y,t) = u(x,y,t)+2v(x,y,t), w(x,y,t)=u(x,y)+v(x,y,t), (x,y)\in P \right\}, \\ 
\izft & = & \text{QP}-\izft\ \ \bigcup \ \text{IP}-\izft.
\eean
\normalsize 
\end{defn}

\begin{note} [Explaining the IZ4$_2$, IZ4$_{10}$ Nomenclature]
The abbreviation IZ4 stands for interleaved $Z_4$-linear.  We use the term $Z_4$-linear to denote that the binary sequence family is derived from a quaternary sequence family that is linear in the quaternary, i.e., $Z_4$ domain.  The reason we have used the word interleaved here is because each sequence in Quaternary Family \calD\ can be expressed as the interleaving of two Family \calA\ ~\cite{Sol,BozHamKum} sequences and this was made extensive use of in the derivations of correlation expressions above.  The subscript $2$ denotes that the family of binary sequences so constructed has period $(2 \times 1023) \ = \ 2046$. The length $1023$ is a fundamental period in GNSS systems on account of the presence of an atomic clock having fundamental frequency 10.23 MHz. To ensure that one period of each spreading code employed is a convenient unit of time, the lengths of the spreading codes are chosen to be a multiple of $1023$.   The spreading code that has been incorporated into the L1 band signal of the Indian navigation satellite system, NavIC, has period $10230$ and in this nomenclature, would correspond to an IZ4$_{10}$ family. 
\end{note}

\begin{note}
We also note that the parameters of Family IZ4$_2$ of size $1024$ are the same as those of the binary family of Generalized Udaya-Siddiqi sequences\cite{TanUdaFan}. The construction is however, different and Family IZ4$_2$ arose here as a by-product of our generation of the modified spreading code family \mfdt. 
\end{note}

\section{Shift-Register Generation}

%

Let
\bean
\beta & = & \alpha(1+2\theta).
\eean
Then $m_{\beta}(x)$ is of the form
\bean
m_{\beta}(x) & = & m_{\alpha}(x) \ + \ 2 g(x), \\
 & = & \big(x^{10}+x^9+x^8+x^6+x^3+x^2+1\big) \ + \ 2 g(x). 
\eean
Setting $x=\beta=\alpha(1+2\theta)$ gives us: 
\bean
m_{\alpha}(\beta) & = & m_{\alpha}(\alpha(1+2\theta)), \\
& = & 2 \theta(\alpha^9 \ + \ \alpha^3), \\
& = & 2 (\alpha^{65} \ + \ \alpha^{64})(\alpha^9 \ + \ \alpha^3), \\
& = & 2 \left\{ (\alpha^{74} \ + \ \alpha^{68})  \ + \ (\alpha^{73} \ + \ \alpha^{67}) \right\}, \\
& = & 2 (\alpha \ + \ 1) \ \ \ \text{as it turns out.} 
\eean
It follows that
\bean
g(x) & = & (x+1),
\eean
and hence that 
\bean
m_{\beta}(x) & = & \big(x^{10}+x^9+x^8+x^6+x^3+x^2+1\big) \ + \ 2 g(x), \\
& = & \big(x^{10}+x^9+x^8+x^6+x^3+x^2+2x+3\big). 
\eean
The parent quaternary sequences have expression of the form 
\bean
Q(x,y,t) & = & x3^t \ + \ T([1+2y]\beta^t), \\
\eean
and hence have characteristic polynomial
\bean
m_{\beta}(x)(x+1) & = & \big(x^{11}+2x^{10}+2x^{9}+x^{8}+x^{7}+x^{6}+x^{4}+2x^{3}+3x^{2}+x+3\big). 
\eean

\subsection{Linear Recursions} 

\scriptsize 
\bean
Q(t+11) & = & 2Q(t+10)+2Q(t+9)+3Q(t+8)+3Q(t+7)+3Q(t+6) \\ & & \ \ \ + \ 3Q(t+4)+2Q(t+3)+Q(t+2)+3Q(t+1)+Q(t).
\eean
Hence if 
\bean
Q(t) & = & U(t) \ + \ 2V(t), 
\eean
it follows that 
\bea \label{eq:linear_recursion} 
U(t+11)  & = &  U(t+8) \ + \ U(t+7) \ + \ U(t+6) \ + \ U(t+4) \ + \ U(t+2) \ + \ U(t+1) \ + \ U(t), \\ \nonumber 
V(t+11) & = & V(t+8) \ + \ V(t+7) \ + \ V(t+6) \ + \ V(t+4) \ + \ V(t+2) \ + \ V(t+1) \ + \ V(t) \\ \nonumber 
&  & + \ U(t+10) \ + \ U(t+9) \ + \ U(t+8) \ + \ U(t+7) \ + \ U(t+6) \ + \ U(t+4) \\ \label{eq:coupled_recursion}
&  & \ + \ U(t+3) \ + \ U(t+1) + \sigma_2\bigg(U(t+8),U(t+7),U(t+6),U(t+4),U(t+2),U(t+1),U(t)\bigg),  
\eea
where the elementary symmetric function is given by 
\bean \scriptsize 
\sigma_2(x_1,x_2,x_3,x_4,x_5,x_6,x_7), & = & 
\prod_{\begin{array}{c} 1 \leq i,j \leq 7\\ j > i  \end{array} } x_i x_j.
\eean
and hence we have that 
by 
\bea \nonumber 
\sigma_2\bigg(U(t+8),U(t+7),U(t+6),U(t+4),U(t+2),U(t+1),U(t)\bigg), \\  \nonumber
= \bigg(U(t+8)+U(t+7)+ U(t+6)+U(t+4)\bigg)\bigg(U(t+2)+U(t+1)+U(t)  \bigg) \\  \nonumber
\ + \  \bigg(U(t+8)+U(t+7)\bigg) \bigg(U(t+6)+U(t+4)\bigg) \ + \ \\  \nonumber
\bigg(U(t+2)+U(t+1)\bigg)U(t) \ + \ U(t+8)U(t+7) \ + \ \\  
+ \ U(t+6)U(t+4) \ + \ U(t+2)U(t+1). 
\eea
This can be implemented using $29$ XOR and $6$ AND gates.   
\normalsize

\begin{figure}
\bc
\includegraphics[width=0.6\textwidth]{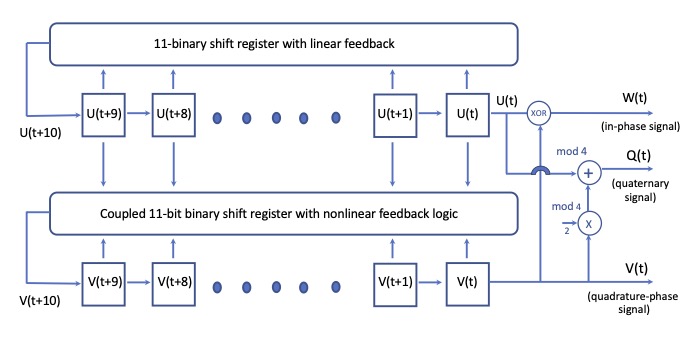} 
\ec
\caption{The IZ4$_2$ spreading code can be generated using $2$ 11-bit binary shift registers.  The upper register is a $7$-tap linear feedback shift register. The lower register is updated using nonlinear logic using the contents of both shift registers. 
The figure also shows where to draw outputs corresponding to generating the in-phase $\{W(t)\}$, quadrature phase sequence $\{V(t)\}$ as well as as composite quaternary spreading code $\{Q(t)\}$. 
}
\label{fig:2K_SR}
\end{figure}

\section{Application to Lunar PNT} 

\subsection{LunaNet Signal-in-Space Document (Aug. 2023 Draft)} 

\begin{figure}\label{fig:lunanet_chipping_rates} 
\bc
\includegraphics[width=4in]{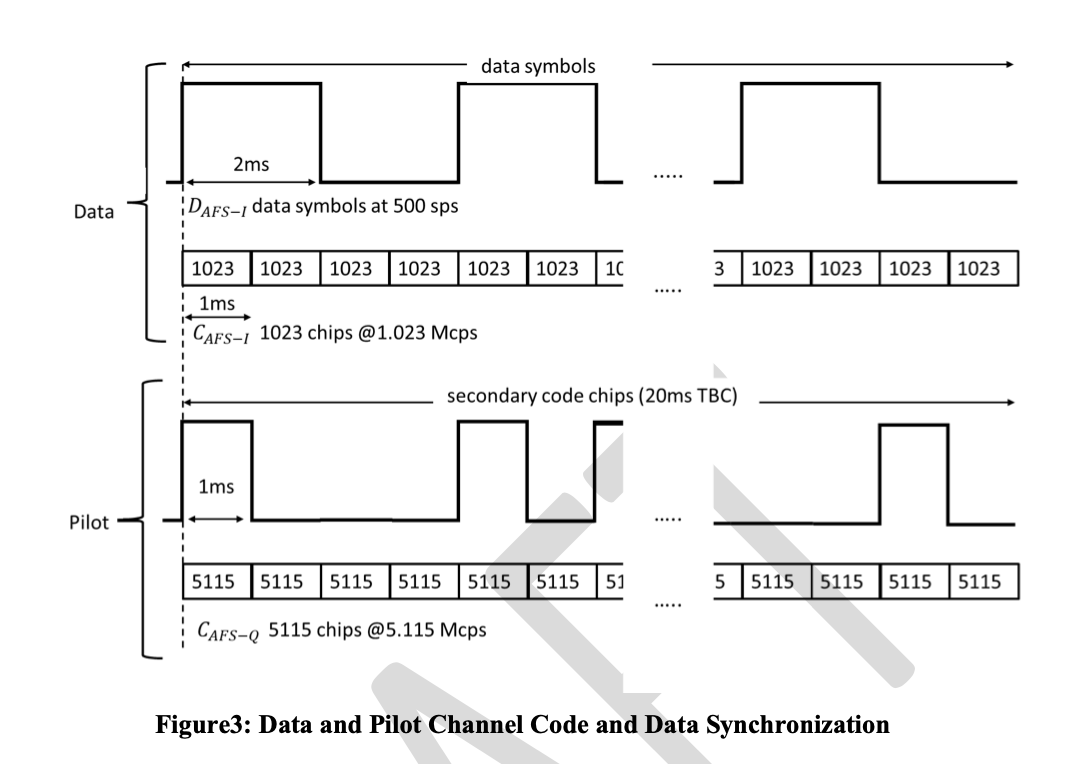} 
\ec
\caption{Data and pilot waveforms, chipping rates and spreading code periods as presented in the LunaNet Signal-in-Space standard\cite{LunaNet_SIS}. }
\end{figure}

Much of the motivation for the material presented in this section, arises from the Draft LunaNet Signal-In-Space Recommended Standard Document Augmented Forward Signal (AFS) (Aug. 2023)\cite{LunaNet_SIS}.  Additional relevant references here include \cite{LunaNet_LNIS,Dafesh1,Dafesh2}.  It has been suggested in \cite{LunaNet_SIS} that spreading codes of length $5115$ (pilot signal) and $1023$ (data signal) are appropriate for a moon-based satellite navigation system, termed the Lunar Augmented Navigation System (LANS), corresponding to chipping rates of $5.115$ MHz and $1.023$MHz respectively (see \cite{LunaNet_SIS}), (see Fig.~\ref{fig:lunanet_chipping_rates}).  

The papers \cite{Dafesh1,Dafesh2} that appeared after the publication of \cite{LunaNet_SIS} have suggested that spreading codes having length $10230$ and $2046$ might offer improved performance and hence be be more appropriate to the setting up of a LANS while maintaining the same chipping rates and the same navigation data rate.  Two spreading codes are proposed in \cite{Dafesh2}.  The proposed spreading code for the pilot signal is the GPS L1CP~\cite{RusISIT,RusNav,Ruspatent} spreading code of length $10230$ employed by GPS (P stands for the pilot signal). The proposed spreading code for the data signal is a truncated Gold sequence (truncated from $2047$) along the lines of the the BDS B1 spreading code of length $2046$ employed by BDS~\cite{BDS}.  

In this part of the paper, we compare the spreading code $\izft$ described in the first part of the paper against the performance of the BDS B1 spreading code~\cite{Dafesh1,BDS}.      

We also compare the performance of the GPS L1C spreading codes as presented in \cite{Dafesh1,Dafesh2} with that of the interleaved $Z_4$-linear spreading code family IZ4$_{10}$.   
Details on the IZ4$_{10}$ spreading code family\footnote{In these references the IZ4$_{10}$ family is referred to simply as the IZ4 family.} can be found in \cite{NavIC_L1_ICD,KumDhaMis_TIT,KumDhaMis_Pat}.

\subsection{Comparing the IZ4$_2$ and BDS B1 Spreading Codes of Length $2046$}

\subsubsection{Tabular Comparison} 

In this table, we make a comparison between the spreading code families IZ4$_2$ and the BDS B1 family.  We also show corresponding parameters of the truncated Gold family presented in \cite{Dafesh2}.   As can be seen the IZ4$_2$ family has significantly lower even-correlation values (in excess of $8$ \dB) and odd-correlation values that are either comparable or else improve upon the corresponding parmeters of the BDS B1 family.  

\bc 
\begin{table} \label{tab:2K}
\caption{Comparing the IZ4$_2$ and the BDS B1 Spreading Code Families} 
\vspace*{0.2in}
\bc 
\begin{tabular}{||c||c||c||c||} \hline \hline \hline 
Property & IZ4$_2$ & Truncated  & BDS B1 \\  
 & & Gold Codes~\cite{Dafesh2} &  \cite{BDS} \\  
\hline \hline \hline
Length & 2046 & 2046 & 2046 \\ \hline  \hline 
Even ACR & -29.83 \dB &   NA & -21.61 \dB\\ \hline 
Even CCR  & -29.83 \dB &   NA & -19.77 \dB \\ \hline 
Odd ACR  & -23.30 \dB &   NA & -22.36 \dB \\ \hline 
Odd CCR  & -20.28 \dB &  -19.6 \dB & -20.28 \dB \\ \hline \hline 
Max (over all even)  & &   &  \\ 
and odd correlations)  & -20.28 \dB &  -19.6 \dB & -20.28 \dB \\ \hline \hline 
RMS (\dB) & -33.11 \dB & -33.1 \dB & -33.11 \dB\\ \hline 
$99\%$ percentile (\dB) & -26.05 \dB & -25.5 \dB  & -25.55 \dB \\ \hline 
$99.9\%$ percentile (\dB) & -23.68 \dB & -23.7 \dB  & -23.68 \dB \\ \hline \hline 
Symbol Balance & $0$ or $2$  &  $0$ & $0$ or $2$ \\ \hline \hline 
Family Size & 221 & 210 & $37$ \\ \hline 
Implementation & \text{SR Based}  & \text{SR based}  & \text{SR based}  \\ \hline \hline 
\end{tabular}
\ec
\end{table}
\ec
\subsubsection{Comparison Based on the CDF} 

The Cumulative Distribution Function (CDF) presents a more complete picture of comparative performance with respect to the correlation metric.  The two CDF plots shown in Fig.~\ref{fig:2K_all_corr} and Fig.~\ref{fig:2K_zoom} below, are plots that include all correlation values even and odd, auto as well as cross-correlation values of the two families \izft\ and the BDS B1 family.  The first of the two plots shows the entire CDF whereas the second plot zooms in onto the portion of the CDF corresponding to values of the CDF $\geq 90\%$, when the CDF is expressed as a percentage. 

\begin{figure}[h!]
\bc
\includegraphics[width=0.8\textwidth]{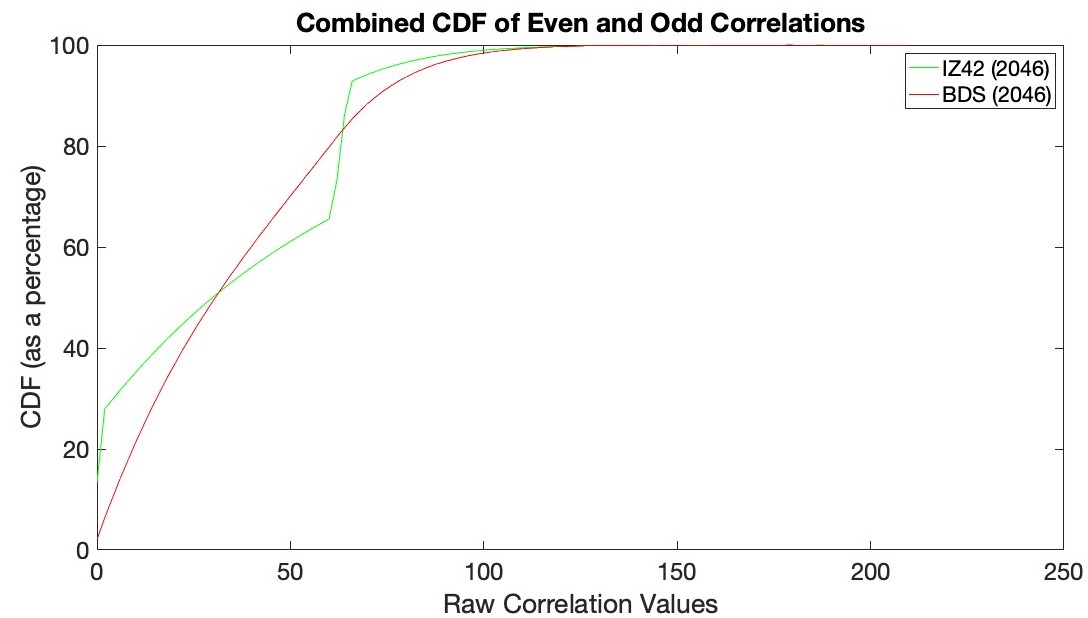} 
\ec
\caption{The figure plots the Cumulative Distribution Function (CDF) of all correlation values of the IZ4$_2$ and BDS B1 (truncated-Gold) spreading codes respectively.  Values of the CDF are expressed as percentages.  } \label{fig:2K_all_corr} 
\end{figure}

\bc
\begin{figure}[h!]\begin{minipage}{4.5in}
\bc
\includegraphics[width=\textwidth]{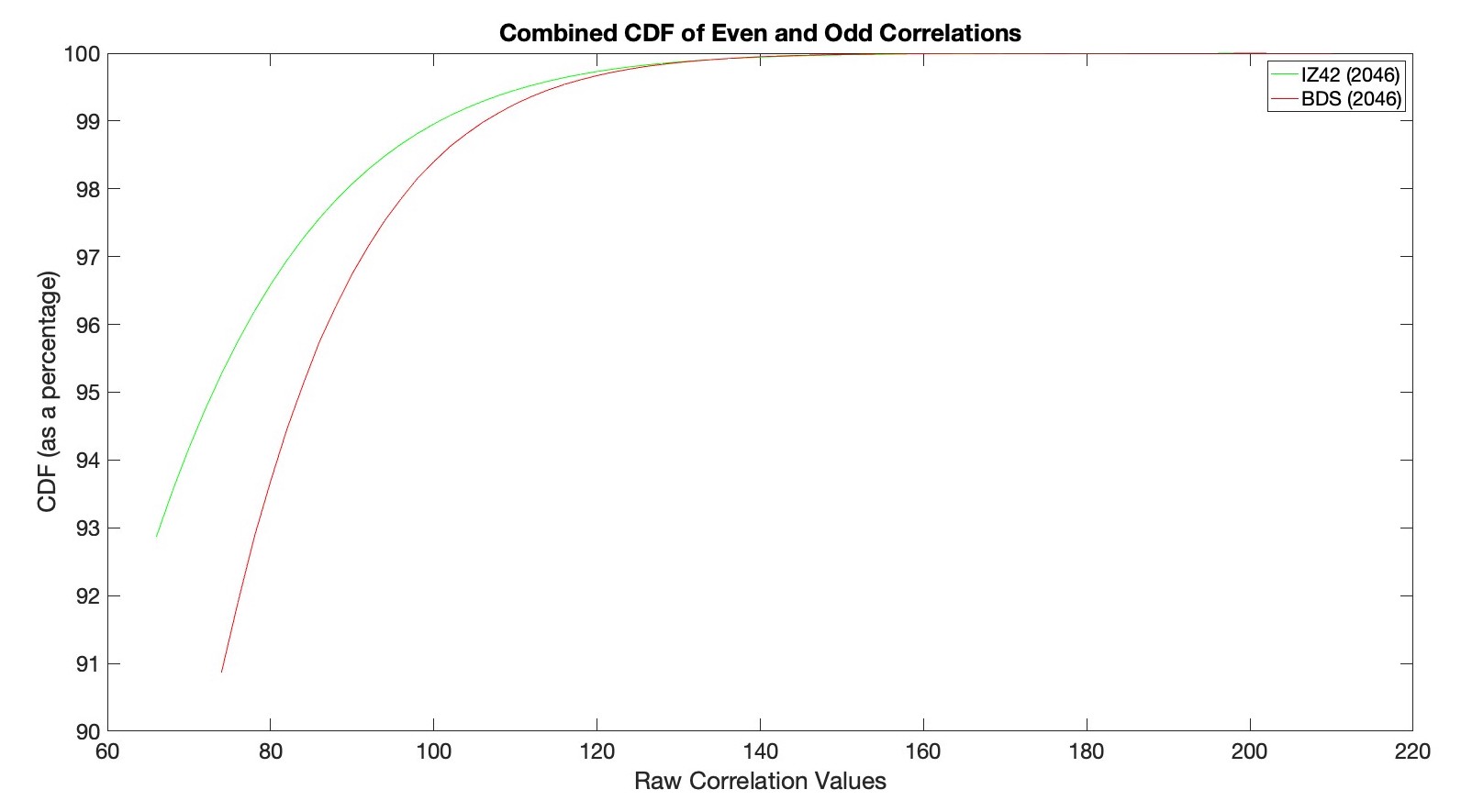} 
\ec
\end{minipage}
\begin{minipage}{2in}
\bc
\includegraphics[width=0.8\textwidth]{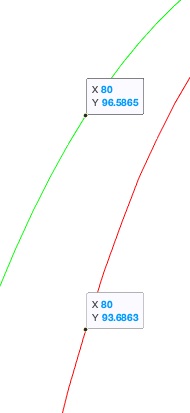} 
\ec
\end{minipage}
\caption{Zooming in to a segment of the  CDF of the correlation values of the IZ4$_2$ and BDS B1 spreading codes.  Only a portion of the CDF corresponding to percentiles $90$ and above is plotted here. As can be seen, above the $95$th percentile, the IZ4$_2$ spreading code does better throughout in terms of having a larger number of correlation values below a specified threshold. As an example, corresponding to a correlation value of $80$ (equivalent to -28.16 \dB), the percentiles associated to the IZ4$_2$ and BDS B1 spreading codes are $96.58$ \% and $93.69$\% respectively.} \label{fig:2K_zoom} 

\end{figure}
\ec

\subsection{Paired Orthogonality}

Although the data and pilot signals are chipped at different rates, one could still regard the data signal as being chipped at the higher speed of $5.115$ MHz except that each symbol remains the same for a duration of $5$ consecutive chips. Under thus interpretations one could ask if it is possible to select pairs of spreading codes for data and pilot in such a way that they are orthogonal at zero cyclic shift (i.e., when time aligned).  In the IZ4$_{10}$ and IZ4$_2$ spreading codes presented here, we have ensured that the first 170 spreading codes from the IZ4$_{10}$ and IZ4$_2$ spreading codes are orthogonal when time aligned. Such a property could potentially be useful in carrier phase tracking. The preference for orthogonality can be found stated for example in column 1, page 3 of \cite{Ruspatent} column 13 of the US Patent \cite{Winkel}.

\subsection{Comparing IZ4$_{10}$ and GPS L1C Ranging Codes (Length $10230$)}

\subsubsection{Tabular Comparison}

In Table~\ref{tab:10K}, we make a comparison between the spreading code families IZ4$_{10}$ and the GPS L1C spreading code families.   As can be seen the IZ4$_2$ family has significantly lower worst-case even-correlation value (in excess of $4.4$ \dB) and odd-correlation values that are either comparable or else improve upon the corresponding parameters of the GPS L1C family.  

\bc 
\begin{table}[h!] \label{tab:10K}
\caption{Comparing the NavIC IZ4$_{10}$ and the GLS L1CP Weil Spreading Code Families} 
\vspace*{0.2in}

\bc
\begin{tabular}{||c||c||c||}  \hline \hline \hline 
Property & IZ4$_{10}$(L1C)  & GPS L1CP  \\  \hline \hline \hline
Length & 10230 & 10230 \\ \hline  \hline 
EACR & -31.7 \dB &   -31.07  \dB \\ \hline 
ECCR  & -31.7 \dB &  -27.21 \dB  \\ \hline 
OACR  & -29.83  \dB &  -28.02 \dB \\ \hline 
OCCR  & -26.5 \dB &  -26.21 \dB \\ \hline \hline 
Max (over all even)  & &     \\ 
and odd correlations)  & -26.5 \dB &  -26.21 \dB  \\ \hline \hline 
RMS (\dB) & -40.11 \dB & -40.11 \dB \\ \hline 
$99\%$ percentile (\dB) &  -32.10 \dB & -32.31 -\dB  \\ \hline 
$99.9\%$ percentile (\dB) &  -30.48 \dB &  -30.26 \dB  \\ \hline \hline 
Symbol Balance & $0$ or $2$  &  0 \\ \hline \hline 
Family Size & 170 & 210 \\ \hline 
Implementation & \text{SR Based}  &   \\ \hline \hline 
\end{tabular}
\ec
\end{table}
\ec

\subsection{Comparison Based on the CDF} 

The Cumulative Distribution Function (CDF) presents a more complete picture of comparative performance with respect to the correlation metric.  The two CDF plots shown in Fig.~\ref{fig:10K_all_corr} and Fig.~\ref{fig:10K_zoom} below, are plots that include all correlation values even and odd, auto as well as cross-correlation values of the two families \izft\ and the BDS B1 family.  The first of the two plots shows the entire CDF whereas the second plot zooms in onto the portion of the CDF corresponding to values of the CDF $\geq 90\%$, when the CDF is expressed as a percentage. 

\begin{figure}[h!] 
\bc
\includegraphics[width=0.8\textwidth]{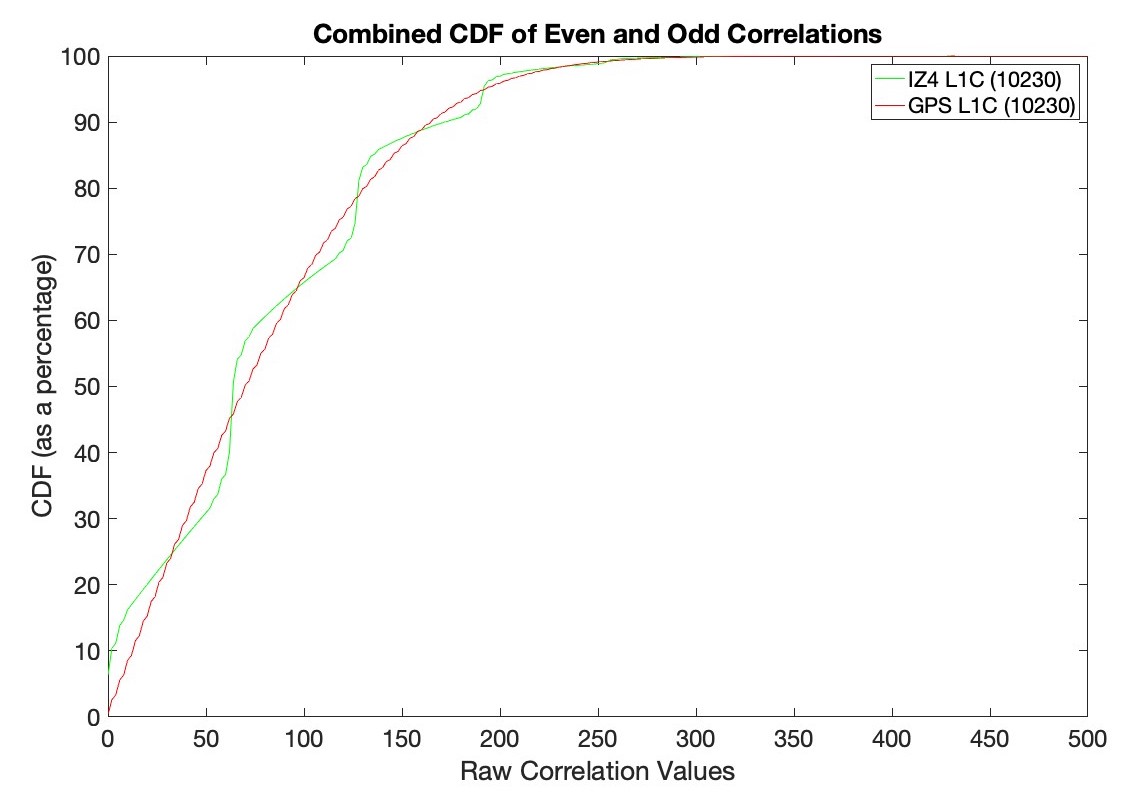} 
\ec
\caption{The figure plots the Cumulative Distribution Function (CDF) of all correlation values of the IZ4$_{10}$ and Weil spreading codes incorporated into ISRO's NavIC L1 signal and the GPS L1 C pilot signal respectively.  Values of the CDF are expressed as percentages.  The maximum correlation magnitudes associated with the IZ4$_{10}$ and the Weil spreading codes are given by $484$ and $500$ respectively. }
\label{fig:10K_all_corr} 
\end{figure}

\bc
\begin{figure}[h!]
\begin{minipage}{5.3in}
\bc
\includegraphics[width=\textwidth]{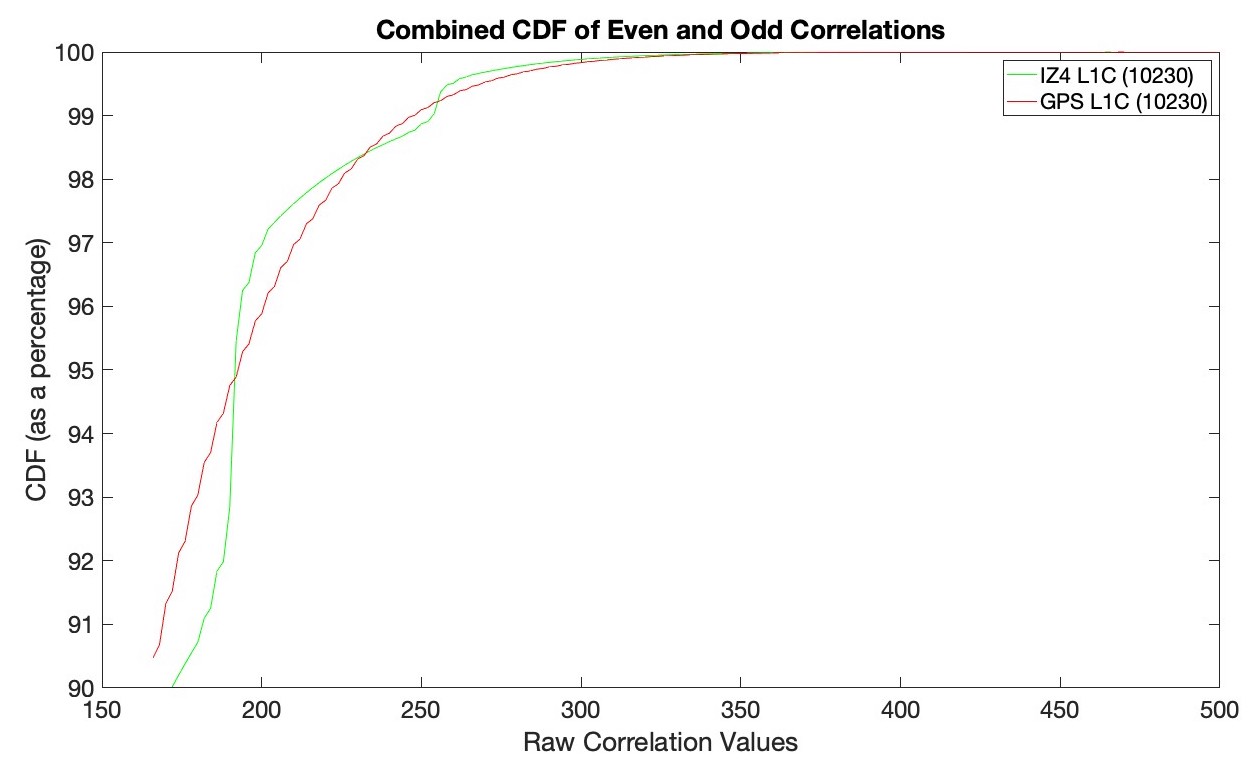} 
\ec
\end{minipage}
\begin{minipage}{1.4in}
\bc
\includegraphics[width=0.95\textwidth]{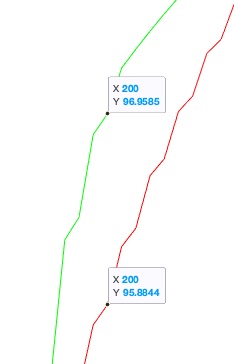} 
\ec
\end{minipage}
\caption{Zooming in to a segment of the  CDF of the correlation values of the IZ4$_{10}$ and Weil spreading codes.  Only a portion of the CDF corresponding to percentiles $90$ and above is plotted here. As can be seen, above the $95$th percentile, apart from a small region, the IZ4$_{10}$ spreading code does better in terms of having a larger number of correlation values below a specified threshold. As an example, corresponding to a correlation value of $200$ (equivalent to -34.18 \dB), the percentiles associated to the IZ4$_{10}$ and Weil spreading codes are $96.96$ \% and $95.88$\% respectively.}
\label{fig:10K_zoom} 
\end{figure}
\ec

\subsection{Ease of Implementation}

\begin{figure}[h!]
\bc
\includegraphics[width=0.6\textwidth]{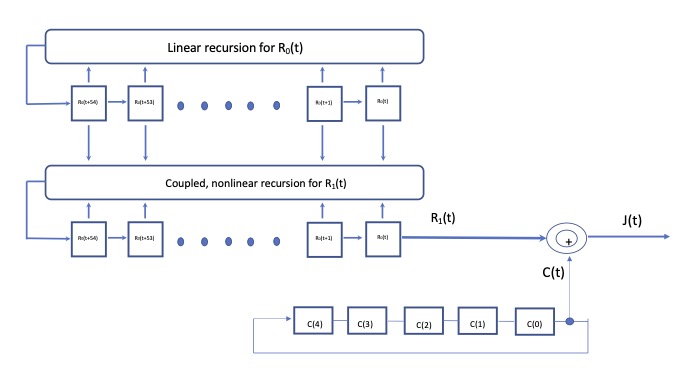} 
\ec
\caption{The IZ4$_{10}$ spreading code can be generated using $2$ 55-bit binary shift registers and an additional $5$-bit pure cycling shift register. The upper register is a linear feedback shift register. The lower register is updated using nonlinear logic using the contents of both shift registers.} 
\label{fig:10K_SR} 
\end{figure}

Ease of implementation is an additional important consideration.  Figure~\ref{fig:2K_SR} above shows that the IZ4$_2$ family affords a simple implementation.  The IZ4$_{10}$ spreading code family also can be implemented using shift registers as shown in Fig.~\ref{fig:10K_SR}. 

\subsection{Some Additional Comments}

\ben
\item The correlation values of the BDS family are for a small family of size $37$.  It is expected that a larger family will suffer degradation in odd correlation properties from what appears in the table above. 
\item The truncated Gold code family described in \cite{Dafesh2} is likely to have similar performance to that of the BDS B1 family. 
\item The IZ4$_{10}$ family is more easily generated than the codes in the GPS L1C Weil code family since the spreading codes in the IZ4 family can be generated using two $55$-bit shift registers and a single $5$-bit shift register. As can be seen from \cite{NavIC_L1_ICD}, a small number of gates suffice to provide the necessary feedback for the two shift registers. \een

\section{Appendix}

\subsection{Proof of Theorem~\ref{thm:closed_form} } \label{sec:closed_form} 

\begin{thm} We have the following closed-form expression for the correlation of two Family \calD\ PRN sequences associated to $y_1,y_2 \in H$ in the case of $\tau$ even:
\bean
\imath^{x_2-x_1}  \phi_{12}(\tau) & = & \bigg( -1 \ - \ (-1)^{(x_1-x_2)} \bigg) \ - \ \\ 
& & 32 \imath^{1-T(e(y_1,y_2,\tau))} \left\{1 \ - \ 
(-1)^{(x_1-x_2)} \bigg(\imath \times (-1)^{tr(e(y_1,y_2,\tau)\theta)}\bigg)\right\}.
\eean
When $\tau$ is odd, we have 
\bean
\imath^{x_2-3x_1}  \phi_{12}(\tau) & = & \bigg( -1 \ - \ (-1)^{(x_1-x_2)} \bigg) \ - \ \\ 
& & 32 \imath^{1-T(e(y_3,y_2,\tau))} \left\{ 1 \ - \ 
(-1)^{(x_1-x_2)} \bigg(\imath \times (-1)^{tr(e(y_3,y_2,\tau)\theta)}\bigg)\right\}, 
\eean
where $y_3=y_1+\theta$. 
\end{thm}
\begin{proof}
In the derivation below, we make use of the following:
\bean
\beta^t & = & \big(\alpha[1+2\theta]\big)^t \ = \  \begin{cases} \alpha^t, & t \text{ even}, \\
\alpha^t(1+2\theta), & t \text{ odd}.\end{cases} 
\eean
Then the cross-correlation $\phi_{12}(\tau)$ of $\{Q_1(t)\}$ and $\{Q_2(t)\}$ is given by: 
\bean
\phi_{12}(\tau) & = & \sum_{t=0}^{2L-1} \imath^{Q_1(t+\tau)-Q_2(t)}.
\eean

\bean
\phi_{12}(\tau) & = & \sum_{t=0}^{2L-1} \imath^{Q_1(t+\tau)-Q_2(t)}, \\
& = & \sum_{t=0}^{2L-1} \imath^{x_13^{t+\tau} \ + \ T([1+2y_1]\beta^{\tau}\beta^t) \ - \ x_2 3^t \ - \ T([1+2y_2]\beta^t)}, \\
& = & \sum_{t \text{ even}} \imath^{(x_13^{\tau} -x_2)\ + \ T([1+2y_1]\beta^{\tau}\beta^t) \ - \ T([1+2y_2]\beta^t)}  \ + \  \sum_{t \text{ odd}}  \imath^{3(x_13^{\tau} -x_2)\ + \ T([1+2y_1]\beta^{\tau}\beta^t) \ - \ T([1+2y_2]\beta^t)}, \\
& = & \imath^{(x_13^{\tau} -x_2)} \bigg( \sum_{t \text{ even}} \imath^{T([1+2y_1]\beta^{\tau}\alpha^t) \ - \ T([1+2y_2]\alpha^t)} \bigg) \ + \  \\ 
& & \imath^{3(x_13^{\tau} -x_2)}  \bigg( \sum_{t \text{ odd}}  \imath^{T([1+2y_1][1+2\theta]\beta^{\tau}\alpha^t) \ - \ T([1+2y_2][1+2\theta]\alpha^t)} \bigg), \\
& = & \imath^{(x_13^{\tau} -x_2)} \bigg(  \sum^{L-1}_{t=0} \imath^{T([1+2y_1]\beta^{\tau}\alpha^t) \ - \ T([1+2y_2]\alpha^t)} \bigg) \ + \  \\
& & \imath^{3(x_13^{\tau} -x_2)}  \bigg( 
\sum_{t=0}^{L-1}  \imath^{T([1+2y_1][1+2\theta]\beta^{\tau}\alpha^t) \ - \ T([1+2y_2][1+2\theta]\alpha^t)}.
\bigg)
\eean

Noting that
\bean
[1+2y_1][1+2\theta] & = & [1+2(y_1+\theta)], 
\eean
we define
\bean
y_3 & = & y_1+\theta, \\
y_4 & = & y_2+\theta. 
\eean
Then we can rewrite the correlation computation as:
\bean
\phi_{12}(\tau) 
& = & \imath^{(x_13^{\tau} -x_2)} \bigg(   \sum^{L-1}_{t=0} \imath^{T([1+2y_1]\beta^{\tau}\alpha^t) \ - \ T([1+2y_2]\alpha^t)} \bigg) \ + \  \\ & & 
 \imath^{3(x_13^{\tau} -x_2)}  \bigg(  \sum_{t=0}^{L-1}  \imath^{T([1+2y_3]\beta^{\tau}\alpha^t) \ - \ T([1+2y_4]\alpha^t)}\bigg).
\eean

\subsubsection{Case $\tau$ Even} 
For $\tau$ even, we can replace $\beta^{\tau}$ by $\alpha^{\tau}$ and we can rewrite the correlation computation as:
\bean
\phi_{12}(\tau) 
& = & \imath^{(x_1 -x_2)} \bigg(   \sum^{L-1}_{t=0} \imath^{T([1+2y_1]\alpha^{\tau}\alpha^t) \ - \ T([1+2y_2]\alpha^t)} \bigg) \ + \  \\ & & 
 \imath^{3(x_1 -x_2)}  \bigg(  \sum_{t=0}^{L-1}  \imath^{T([1+2y_3]\alpha^{\tau}\alpha^t) \ - \ T([1+2y_4]\alpha^t)}\bigg), \\
& & \\ 
& = & \imath^{(x_1 -x_2)} \underbrace{\psi(y_1,y_2,\tau)}_{\text{Fam ${\cal A}$ correlation}}  \ + \  \imath^{3(x_1 -x_2)}  \underbrace{\psi(y_3,y_4,\tau)}_{\text{Fam ${\cal A}$ correlation}},\\ 
& = & \imath^{(x_1 -x_2)} \bigg( \underbrace{\psi(y_1,y_2,\tau)}_{\text{Fam ${\cal A}$ correlation}}  \ + \  (-1)^{(x_1 -x_2)}  \underbrace{\psi(y_3,y_4,\tau)}_{\text{Fam ${\cal A}$ correlation}}\bigg),
\eean
where
\bean
\psi(y_1,y_2,\tau) & := & \sum^{L-1}_{t=0} \imath^{T([1+2y_1]\alpha^{\tau}\alpha^t) \ - \ T([1+2y_2]\alpha^t)}, \\
\psi(y_3,y_4,\tau) & := & \sum_{t=0}^{L-1}  \imath^{T([1+2y_3]\alpha^{\tau}\alpha^t) \ - \ T([1+2y_4]\alpha^t)}.
\eean
\begin{note}
Note that the correlation of two quaternary sequences belonging to Family ${\cal A}$~\cite{Sol,BozHamKum} correlation value $\psi(y_1,y_2,\tau)$ is only a function of $\tau \pmod{L}$. 
\end{note}

\subsubsection{Case $\tau$ Odd} 
For $\tau$ odd, we can replace $3^{\tau}$ by $3$ and $\beta^{\tau}$ by $\alpha^{\tau}(1+2\theta)$.  
In comparison to the case of $\tau$ even, this has the impact of replacing:
\bean
& \begin{cases} x_1 \ \text{by } \ 3x_1, \\
x_2 \ \text{remains } \ x_2, \\ 
y_1 \ \text{ by } \ y_1+\theta, \\
y_2 \ \text{ remains } \ y_2, \\
y_3 \ \text{ by } \ y_3+\theta \ = \ y_1, \\
y_4 \ \text{ remains } \ y_4. 
\end{cases} 
\eean
It follows that:
\bean
\phi_{12}(\tau) 
& = & \imath^{(3x_1 -x_2)} \bigg( \underbrace{\psi(y_3,y_2,\tau)}_{\text{Fam ${\cal A}$ correlation}}  \ + \  (-1)^{(3x_1 -x_2)}  \underbrace{\psi(y_1,y_4,\tau)}_{\text{Fam ${\cal A}$ correlation}}\bigg), \\
& = & \imath^{(3x_1 -x_2)} \bigg( \underbrace{\psi(y_3,y_2,\tau)}_{\text{Fam ${\cal A}$ correlation}}  \ + \  (-1)^{(x_1 -x_2)}  \underbrace{\psi(y_1,y_4,\tau)}_{\text{Fam ${\cal A}$ correlation}}\bigg).  
\eean

\subsubsection{Correlation Expressions for $\phi_{12}(\tau)$ in Summary}
We have in summary:
\bea \label{eq:Bone_qc} 
\phi_{12}(\tau) 
& = & \begin{cases} \imath^{(x_1 -x_2)} \bigg( \psi(y_1,y_2,\tau)  \ + \  (-1)^{(x_1 -x_2)}  \psi(y_3,y_4,\tau)\bigg), \text{ for $\tau$ even  }, \\ 
\imath^{(3x_1 -x_2)} \bigg( \psi(y_3,y_2,\tau)  \ + \  (-1)^{(x_1 -x_2)}  \psi(y_1,y_4,\tau)\bigg), \text{ for $\tau$ odd },
\end{cases}  
\eea
where
\bean
y_3 & = & (y_1+\theta), \\
y_4 & = & (y_2+\theta), 
\eean
and
\bean
\psi(y_1,y_2,\tau) & := & \sum^{L-1}_{t=0} \imath^{T([1+2y_1]\alpha^{t+\tau}) \ - \ T([1+2y_2]\alpha^t)}, 
\eean
etc.

\subsubsection{Can $|\phi_{12}(\tau)|$ be Large of $O(L)$ ?}
From \eqref{eq:Bone_qc}, we see that this can happen iff one of the following holds: 
\bean
y_1=y_2 & \text{ and } & \tau \ = 0,  \\
y_3=y_4 \ \ \Leftrightarrow \ y_1=y_2  & \text{ and } & \tau \ =\ 0,   \\
y_3=y_2 \ \ \Leftrightarrow \ y_1+\theta=y_2  & \text{ and } & \tau \ = \ L, \\ 
y_1=y_4 \ \ \Leftrightarrow \ y_1=y_2+\theta.  & \text{ and }  & \tau \ = \ L.  
\eean
However,
\bean
y_1=y_2 & \text{ and } & \tau \ = 0,  \Rightarrow \phi_{12}(\tau)=0, \text{ as } x_1 \neq x_2, \\
y_3=y_4 \ \ \Leftrightarrow \ y_1=y_2  & \text{ and } & \tau \ =\ 0,  \Rightarrow \phi_{12}(\tau)=0, \text{ as } x_1 \neq x_2.
\eean
We safeguard the remaining two settings by ensuring that $y \in H \Rightarrow (y+\theta) \not \in H$: 
\bean
y_3=y_2 \ \ \Leftrightarrow \ y_1+\theta=y_2  & \text{ and } & \tau \ = \ L, \\ 
y_1=y_4 \ \ \Leftrightarrow \ y_1=y_2+\theta.  & \text{ and }  & \tau \ = \ L. 
\eean
Thus to ensure good quaternary correlation, we impose the condition that 
\bean
y \in H & \Rightarrow & (y+\theta) \not \in H. 
\eean

\begin{defn} Let $H \subseteq \ffm$ be such that \bean
y \in H & \Rightarrow & (y+\theta) \not \in H. 
\eean
Let Family \calD\ \cite{TanUda} be defined by 
\bean
\calD & = & \left\{ \{x3^t \ + \ T([1+2y] \beta^t)\} \mid  x \in \{0,1\}, \ y \in H \right\}. 
\eean
Note that this family has size $|\calD | \ = \ (2 \times 512) \ = \ 1024$. 
\end{defn}

\subsubsection{Closed-Form Expressions}

\subsubsection{Quaternary Correlation Values of Family \calA} 

We know from the analysis in \cite{KumDhaMis_TIT} , that 
\bean
\psi(y_1,y_2,\tau) & = & -1 - 32\imath^{1-T(e(y_1,y_2,\tau))},
\eean
where
\bean
e(y_1,y_2,\tau) & \in & {\cal T}_m \ \ \text{ (Teichmuller set)}, \\
e(y_1,y_2,\tau) & = & \mu \ + \ y_1 \ + \ (y_1+y_2)\mu^2, \pmod{2}\\
\mu & = & \sqrt{\frac{1}{1+\alpha^{\tau}}}.
\eean
\begin{note}
Since $e(y_1,y_2,\tau)$ belongs to ${\cal T}_m$ its value is uniquely defined by its value modulo $2$. 
\end{note}

\subsubsection{Case When $\tau$ is Even}

It follows then that for $\tau$ even: 
\bean
\phi_{12}(\tau) & = & \imath^{x_1-x_2} \left\{ \bigg(-1 - 32\imath^{1-T(e(y_1,y_2,\tau))}\bigg) \ + \ 
(-1)^{(x_1-x_2)} \bigg(-1 - 32\imath^{1-T(e(y_3,y_4,\tau))}\bigg)\right\},
\eean
i.e.,
\bean
\imath^{x_2-x_1} \phi_{12}(\tau) & = & \left\{ \bigg(-1 - 32\imath^{1-T(e(y_1,y_2,\tau))}\bigg) \ + \ 
(-1)^{(x_1-x_2)} \bigg(-1 - 32\imath^{1-T(e(y_3,y_4,\tau))}\bigg)\right\},
\eean

Next note that 
\bean
e(y_3,y_4,\tau) & = & \mu \ + \ y_3 \ + \ (y_3+y_4)\mu^2, \pmod{2} \\
& = & e(y_1,y_2,\tau) \ + \ \theta, \pmod{2}. 
\eean
Thus we have
\bean
e(y_3,y_4,\tau) & = & \reallywidehat{e(y_1,y_2,\tau) \ + \ \theta},
\eean
where
\bean
\reallywidehat{e(y_1,y_2,\tau) \ + \ \theta} & = & e(y_1,y_2,\tau) \ + \ \theta \ + \ 2\sqrt{\theta e(y_1,y_2,\tau)}.
\eean

As a result, we can write: {\scriptsize 
\bean 
\imath^{x_2-x_1} \phi_{12}(\tau) & = & \bigg( -1 \ - \ (-1)^{(x_1-x_2)} \bigg) \ + \ \\ 
& & \left\{ \bigg(- 32\imath^{1-T(e(y_1,y_2,\tau))}\bigg) \ + \ 
(-1)^{(x_1-x_2)} \bigg(- 32\imath^{1-T(e(y_3,y_4,\tau))}\bigg)\right\}, \\
& = & \bigg( -1 \ - \ (-1)^{(x_1-x_2)} \bigg) \ - \ \\ 
& & 32 \left\{ \bigg(\imath^{1-T(e(y_1,y_2,\tau))}\bigg) \ + \ 
(-1)^{(x_1-x_2)} \bigg(\imath^{1-T(e(y_3,y_4,\tau))}\bigg)\right\}, \\
& = & \bigg( -1 \ - \ (-1)^{(x_1-x_2)} \bigg) \ - \ \\ 
& & 32 \imath^{1-T(e(y_1,y_2,\tau))} \left\{ 1 \ + \ 
(-1)^{(x_1-x_2)} \bigg(\imath^{-T(\theta)-2T(e(y_1,y_2,\tau)\theta)}\bigg)\right\}, \\
& = & \bigg( -1 \ - \ (-1)^{(x_1-x_2)} \bigg) \ - \ \\ 
& & 32 \imath^{1-T(e(y_1,y_2,\tau))} \left\{ 1 \ - \ 
(-1)^{(x_1-x_2)} \bigg(\imath \times (-1)^{tr(e(y_1,y_2,\tau)\theta)}\bigg)\right\}, 
\eean
}

since by our choice of $\theta \ = \ (\alpha^{64}+\alpha^{65}) \pmod{2}$ we have that $T(\theta)=1$. 

\end{proof}

\subsection{Component Binary Correlations in Terms of Quaternary Correlations} 
\label{sec:bcorr}

\begin{thm}
\bean
\rho_v(\tau) & = & \Re\big(\phi_{12}(\tau)\big) \ + \ \Im\big(\Delta_{12}(\tau)\big), \\ 
\rho_w(\tau) & = & \Re\big(\phi_{12}(\tau)\big) \ - \ \Im\big(\Delta_{12}(\tau)\big), \\ 
\rho_{vw}(\tau) & = & \Re\big(\Delta_{12}(\tau)\big) \ + \ \Im\big(\phi_{12}(\tau)\big).
\eean
where
\bean
\phi_{12}(\tau) & = & \sum_{t=0}^{2L-1} C_1(t+\tau)C^*_2(t), \ \ \ \text{(Family ${\cal D}$ correlation)}\\
\Delta_{12}(\tau) & = & \sum_{t=0}^{2L-1} C_1(t+\tau)C_2(t), \ \ \ \text{(we will term this as a Family \calD\ anti-correlation)}. 
\eean
\end{thm}

\begin{proof} 

\subsubsection{Quadrature-Phase Component Correlation Correlation}

\bean
\rho_v(\tau) & = & \sum_{t=0}^{2L-1} (-1)^{v_1(t+\tau)-v_2(t)}, \\
& = & \sum_{t=0}^{2L-1} \big( \delta^* C_1(t+\tau) \ + \ \delta C^*_1(t+\tau) \big) \times \big( \delta^* C_2(t) \ + \ \delta C^*_2(t) \big), \\
& = & (\delta^*)^2 \sum_{t=0}^{2L-1} C_1(t+\tau)C_2(t) \ + \ 
|\delta|^2 \sum_{t=0}^{2L-1} C^*_1(t+\tau)C_2(t) , \\
& & \ \ \ + \ |\delta|^2 \sum_{t=0}^{2L-1} C_1(t+\tau)C^*_2(t) \ + \ 
(\delta)^2 \sum_{t=0}^{2L-1} C^*_1(t+\tau)C^*_2(t) , \\
& = & (\delta^*)^2 \Delta_{12}(\tau) \ + \ 
|\delta|^2 \phi^*(\tau)  \ + \ |\delta|^2 \phi_{12}(\tau) \ + \ 
(\delta)^2 \Delta^*(\tau), \\
& = & \frac{1}{2}\big( \phi_{12}(\tau)+\phi^*(\tau)\big) \ + \ \frac{i}{2}
\big( \Delta^*(\tau)-\Delta_{12}(\tau)\big), \\
& = & \Re\big(\phi_{12}(\tau)\big) \ + \ \Im\big(\Delta_{12}(\tau)\big) ,
\eean
where
\bean
\phi_{12}(\tau) & = & \sum_{t=0}^{2L-1} C_1(t+\tau)C^*_2(t), \ \ \ \text{(Family ${\cal D}$ correlation)}\\
\Delta_{12}(\tau) & = & \sum_{t=0}^{2L-1} C_1(t+\tau)C_2(t), \ \ \ \text{(we will term this as a Family \calD\ anti-correlation)}. 
\eean

\subsubsection{In-Phase Component Correlation Computation} 

\bean
\rho_w(\tau) & = & \sum_{t=0}^{2L-1} (-1)^{w_1(t+\tau)-w_2(t)}, \\
& = & \sum_{t=0}^{2L-1} \big( \delta C_1(t+\tau) \ + \ \delta^* C^*_1(t+\tau) \big) \times \big( \delta C_2(t) \ + \ \delta^* C^*_2(t) \big), \\
& = &\delta^2 \Delta_{12}(\tau) \ + \ (\delta^*)^2 \Delta^*(\tau) \ + \ 
|\delta|^2 \bigg(  \phi_{12}(\tau) \ + \ \phi^*(\tau) \bigg) , \\
& = & \Re\{\phi_{12}(\tau)\} \  - \ \Im\{\Delta_{12}(\tau)\}. 
\eean
where
\bean
\phi_{12}(\tau) & = & \sum_{t=0}^{2L-1} C_1(t+\tau)C^*_2(t), \ \ \ \text{(Family \calD\ correlation)}\\
\Delta_{12}(\tau) & = & \sum_{t=0}^{2L-1} C_1(t+\tau)C_2(t), \ \ \ \text{(Family \calD\  anti-correlation)}. 
\eean

\subsubsection{Cross-Phase Component Correlation Computation} 

\bean
\rho_{v,w}(\tau) & = & \sum_{t=0}^{2L-1} (-1)^{v_1(t+\tau)-w_2(t)}, \\
& = & \sum_{t=0}^{2L-1} \big( \delta^* C_1(t+\tau) \ + \ \delta C^*_1(t+\tau) \big) \times \big( \delta C_2(t) \ + \ \delta^* C^*_2(t) \big), \\
& = & |\delta|^2 \bigg(\Delta_{12}(\tau) \ + \ \Delta^*(\tau)\bigg) \ + \ \delta^2 \phi^*(\tau) \ + \
 (\delta^*)^2 \phi_{12}(\tau), \\
 & = & \Re\{\Delta_{12}(\tau)\} \ + \ \Im\{\phi_{12}(\tau)\}. 
\eean
where
\bean
\phi_{12}(\tau) & = & \sum_{t=0}^{2L-1} C_1(t+\tau)C^*_2(t), \ \ \ \text{(Family \calD\ correlation)}\\
\Delta_{12}(\tau) & = & \sum_{t=0}^{2L-1} C_1(t+\tau)C_2(t), \ \ \ \text{(Family \calD\  anti-correlation)}. 
\eean

\end{proof}

\section*{Acknowledgement}

We would like to acknowledge the support received from distinguished scientists and senior administrators within ISRO as well as senior faculty and administrators at IISc.

\end{document}